# Conditional Handover Modelling for Increased Contention Free Resource Use in 5G-Advanced

Jędrzej Stańczak[1,3], Umur Karabulut[2] and Ahmad Awada[2]
[1]Nokia Standards, Poland, [2]Nokia Standards, Germany, [3]Wroclaw University of Science and Technology, Poland
E-mail: jedrzej.stanczak@nokia.com, umur.karabulut@nokia.com, ahmad.awada@nokia.com

*Abstract*—This paper elaborates on Conditional Handover (CHO) modelling, aimed at maximizing the use of contention free random access (CFRA) during mobility. This is a desirable behavior as CFRA increases the chance of fast and successful handover. In CHO this may be especially challenging as the time between the preparation and the actual cell change can be significantly longer in comparison to non-conditional handover. Thus, new means to mitigate this issue need to be defined. We present the scheme where beam-specific measurement reporting can lead to CFRA resource updating prior to CHO execution. We have run system level simulations to confirm that the proposed solution increases the ratio of CFRA attempts during CHO. In the best-case scenario, we observe a gain exceeding 13%. We also show how the average delay of completing the handover is reduced. To provide the entire perspective, we present at what expense these gains can be achieved by analyzing the increased signaling overhead for updating the random access resources. The study has been conducted for various network settings and considering higher frequency ranges at which the user communicates with the network. Finally, we provide an outlook on future extensions of the investigated solution.

## I. INTRODUCTION

Conditional Handover (CHO) is an important part of New Radio (NR) mobility framework. It has been standardized in 3GPP Release 16 to enhance the reliability and robustness of handover in cellular networks (NWs). Since then, it has been adopted by multiple other technologies, such as Non-Terrestrial Networks (NTN) or Integrated Access Backhaul (IAB). The performance of CHO in NTN has been investigated in [1], while a general description of CHO and how it improves mobility at higher frequency ranges can be found in [2] and [3], respectively. CHO is also subject to various further research studies, such as on enabling fast CHO (FCHO) by keeping the target cell configurations even after handover execution. As claimed in [4], this can be especially useful for frequency range 2 (FR2), at the problematic cell boundaries where subsequent cell change may be initiated rapidly upon completing the previous handover.

The benefits of CHO predominantly stem from the separation of handover preparation and execution phases. The preparation occurs when the radio link between the User Equipment (UE) and the Base Station (BS) is still of sufficient quality, so the risk of failure in receiving the handover configuration is reduced. The actual cell change happens only if the NW-configured, UE-evaluated condition is met. As can be inferred from the aforementioned CHO principle, when executing handover, the UE applies a configuration provided by the source cell during the preparation phase. It implies the configuration is given to the UE early and it may become at least partly suboptimal (e.g. regarding the beam-specific resource assignments) at the time of cell change (i.e. CHO execution). According to the study in [5], the time between CHO preparation and execution can be as large as 9-10 seconds (cf. baseline handover where the execution happens immediately upon the reception of the handover command). This period is long enough to invalidate the CFRA-related CHO preparations due to UE movement or changes in the radio propagations, especially at FR2.

Random Access (RA) preambles are the key components of the configuration used by the UE at the time of handover. In NR these can be configured per individual beam (see *RACH-Config* in [6]). In particular at higher frequencies (e.g. in FR2) the signal quality or level can be subject to abrupt changes. Thus, if it drops significantly, then the entire RA may fail when performed on the beam configured with CFRA resources whose quality has deteriorated.

The main difference between Contention Based Random Access (CBRA) and CFRA which is relevant to this work is that in the former, there are no dedicated, UE-specific preambles and the risk of collisions during mobility is higher, whereas in the latter, the preambles for RA are pre-assigned per UE. Thus, the possibility of seamless and fast handover increases if CFRA is applied. However, CFRA resources are limited and therefore shall be allocated in a well-considered manner – also by allowing to use them only if the radio conditions are sufficiently favourable.

RA and associated resource reservation have been studied in multiple research papers. The authors of [7] propose a method of reserving RA resources specifically for Ultra-Reliable Low Latency Communication (URLLC) to meet stringent delay requirements. In [8] it is analysed how to reduce the complexity and latency of RA by adopting two-step approach, available in 3GPP specification since Release 16. While two-step RA allows the reduction of the overall time required to complete RA, it can be used predominantly in cases where accurate estimation of timing advance (TA) is not essential.

The delay and resource reservation-related aspects of CFRA and CBRA are further elaborated also in the next sections of this paper whose general aim is to describe the RA challenges when CHO is applied. We also propose methods how the identified issue can be mitigated. This type of analysis and dedicated solution is not available in the investigated state of the art which is focused on the overall aspects related to RA in NR, rather than on optimizing CFRA during CHO.

The paper is organized as follows. Section II comprehensively describes RA, the issue of outdated CFRA resources in CHO and proposes how to mitigate it. In Section III the simulation environment is introduced, analysed metrics are explained and performance results are shown and interpreted. Finally, Section IV reiterates the main findings and discusses potential next steps.

## II. UPDATING CFRA RESOURCES IN CHO

### A. Random Access During Handover

Random Access is an essential procedure used in cellular networks during the initial access, connection re-establishment, radio link failure recovery and many more [9]. Its efficient completion is also of utmost importance during cell change. To minimize the interruption time during handover, defined as the period throughout which the UE cannot exchange data with the BS, RA must be rapidly completed so that the UE is able to send or receive the data via network scheduled transmissions.

Detailed message sequences of CBRA and CFRA are shown in Fig. 1 and Fig. 2, respectively. In CBRA the UE sends RA Preamble which is chosen randomly from the pool shared with other UEs served by the BS. Thus, there is a risk of collisions since more than one UE can randomly select the same CBRA preamble and use it simultaneously. In that case, the contention may not be resolved, unlike shown in the final step in Fig. 1, as more than a single UE will use the same UL grant for scheduled transmission in step 3 of Fig. 1. As a result, the entire CBRA procedure must be repeated (i.e. the UE reselects the RA preamble and sends it to the BS, while another collision is also possible). The principle of CFRA is to resolve the aforementioned issue by applying UE-specific preamble assignment (first step in Fig. 2). This allows the UE to use a dedicated set of resources when the actual RA is performed and eliminates the risk of contention with other users. That in turn should increase the chance of successful HO, accelerate the entire mobility procedure and reduce the period with no data exchange between the UE and the BS.

The delay introduced by RA has been evaluated for each cellular generation, most recently when NR was designed (i.e. 3GPP Release 15). Table I presents RA components with their minimum and average values for the subcarrier spacing (SCS) of 15 kHz, typical for Long Term Evolution (LTE) and NR deployments. The numbers provided are compliant with [10]. The first three components are applicable to CFRA, while all elements of Table I are needed to declare CBRA as completed. As can be estimated by adding the delay of relevant components in Table I, the minimum delay ($D_{CFRA}^{min}$) for CFRA is 4.5 ms, while the average delay ($D_{CFRA}^{avg}$) is equal to 8.5 ms. For CBRA these values are substantially longer and equal to 15.5 ms ($D_{CBRA}^{min}$) and 19.5 ms ($D_{CBRA}^{avg}$). Similar numbers for four-step and two-step RA have been confirmed via simulation study, as presented in [8].

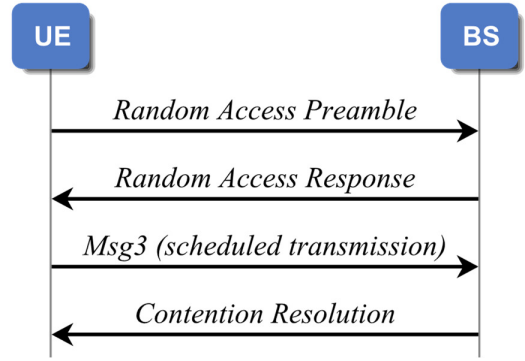

Fig. 1. Contention-Based Random Access (CBRA).

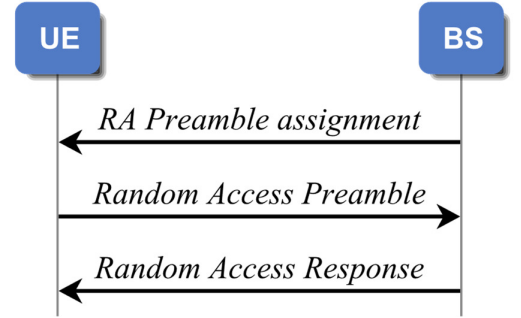

Fig. 2. Contention-Free Random Access (CFRA).

TABLE I: RANDOM ACCESS LATENCY COMPONENTS

| | Component | Minimum Delay [ms] | Average Delay [ms] |
|---|---|---|---|
| 1 | RA scheduling period | 0.5 | 2.5 |
| 2 | Transmission of Random Access preamble | 1 | 1 |
| 3 | RA preamble detection and transmission of RAR | 3 | 5 |
| 4 | UE processing delay (for uplink grant) | 5 | 5 |
| 5 | Transmission of Msg3 (RRC message) | 1 | 1 |
| 6 | BS processing delay | 4 | 4 |
| 7 | Transmission of Contention Resolution | 1 | 1 |

Dedicated preambles for handover related CFRA are provided as a part of Handover Command (i.e. Radio Resource Control Reconfiguration message [6], sent from the source BS to configure the UE for mobility). They are assigned per individual beam and configured to be used only during a specific period, denoted as *RA occasion*.

If the UE measures a candidate target cell $c_i$ ($c_i \epsilon C$, $C = \{c_1, ..., c_M\}$, where M is the maximum number of prepared CHO candidate cells per each UE) with the set of downlink beams $B$ whose cardinality is equal to $N^{beams}$ beams per cell $c_i$, the UE will have CFRA resources assigned by the BS just for the subset $B^{CFRA}$, where the number of elements in $B^{CFRA}$ is usually smaller than $N^{beams}$ defining the size of $B$.

The size of $B^{CFRA}$ is limited as dedicated preambles are scarce and need to be wisely distributed, as we have explained above. This implies the BS needs to predict which beams will be most suitable for accessing each target cell at the time of HO execution ($t_{HO}$), i.e. will fulfil the following condition

$$P_{b_j,c_i}^{CFRA}(t_{HO}) \geq Thr_{CFRA} \quad (1)$$

where $P_{b_j,c_i}^{CFRA}$ is the beam-level Reference Signal Received Power (RSRP) measured for downlink beam $b_j$ of cell $c_i$ at time $t_{HO}$, while $Thr_{CFRA}$ is the threshold configured by the BS and used to decide if CFRA resources for $b_j$ can be used. Measurement reports (MR) sent by the UE to the BS are used for such predictions done by the network. This assessment may be relatively easy for non-conditional HO, whereas it is challenging for CHO (as underlined above, the handover execution and preparation phases are decoupled). Due to these reasons, new means for making CFRA resources useful, also in case of CHO, are needed. In the subsequent section we describe how the issue can be mitigated.

### B. Method for Updating CFRA During Conditional Handover

To ensure the UE has CFRA resources assigned to the beams $b_j$ which are sufficiently good at the time of CHO execution, i.e. fulfil inequality (1), a new method is proposed, wherein the UE can receive an updated configuration comprising just CFRA resources. This occurs after CHO preparation and before CHO execution. In order to provide a new subset of beams configured with dedicated CFRA resources ($B^{CFRA}$), the source BS needs to receive a MR from the UE. The MR is used to decide for which beams the resources shall be updated. The decision is not entirely up to the source BS. CFRA resources are assigned from the pool managed by the target BS, as they concern the beams belonging to target cell $c_i$. Thus, it is the latter entity that allows to associate new beams with CFRA resources. The principles of this method are depicted in Fig. 3. The first four steps are compliant with 3GPP Release 16 CHO [9]. Novel part starts afterwards, from step 5, wherein the UE sends a MR if it observes that the measured signal level (e.g., RSRP) for any of the beams prepared with CFRA resources ($P_{b_j,c_i}^{CFRA}$) is worse by at least an offset than the signal level measured for the non-prepared beam of the same candidate cell $c_i$ ($P_{b_k,c_i}^{Non-CFRA}$), as per the following inequality

$$P_{b_k,c_i}^{Non-CFRA}(t) \geq P_{b_j,c_i}^{CFRA}(t) + offset, b_k \neq b_j, b_k, b_j \in B \quad (2)$$

This implies those CFRA resources are likely to become unusable when CHO execution happens, so CFRA resource assignment to the currently better beam is desirable if the ratio of contention-free access should be maximized. To perform this, a corresponding request is sent from the source BS to target BS (step 6) and new resources are provided if target BS acknowledges it (step 7). In step 8 those new assignments are communicated to the UE and from this point in time, if CHO execution is triggered (step 9), it will rely on CFRA resources for updated beams.

CHO framework, as specified in 3GPP Release 16, lacks such beam-based MR triggering and resulting CFRA resources reassignment. The issue of suboptimal resources for RA in CHO was raised in [11], but at the time of writing, this has not been addressed in standardization.

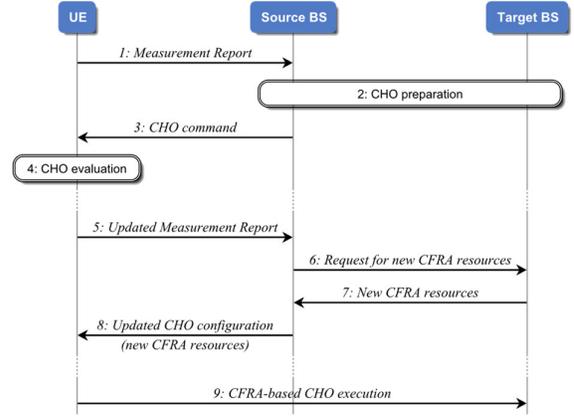

Fig. 3. CFRA resource updating prior to CHO execution.

In the following section we present the simulation analysis which shows how the scheme described above improves the conditional mobility in NR thanks to the increased use of CFRA.

## III. PERFORMANCE EVALUATION

This section presents the performance of CFRA updating procedure which has been introduced earlier in this paper. We begin with the description of the scenario, parameters and investigated KPIs. Then the results of the simulation study are depicted and analysed.

### A. Simulation Scenario and Parameters

In Table II key system-level simulation parameters are shown. The study has been conducted using proprietary MATLAB-based tool in Urban-Micro (UMi) deployment scenario. Seven BSs were used, each with three hexagonal cells. The carrier frequency is from FR2 range and equal to 28 GHz. The UEs have been distributed randomly and moved with constant velocity.

The time needed to update CFRA resources ($T_{update}$) is equal to the joint duration of steps 5 – 8 in Fig. 3. It includes twice the Uu interface (connecting the UE and Source BS) delay, twice the Xn interface (connecting Source and Target BS) delay plus the required processing time at all involved entities. The value used in the simulations for $T_{update}$ (30 ms) implies that in some cases the CFRA resource updating was not effectively completed if during $T_{update}$ the UE has triggered the CHO. This is due to the principle that handover execution should not be suspended if the associated condition has been met, even if beam-specific resources for RA are not updated.

The preparation of CHO is initiated when a candidate cell is worse by 10 dB, 3 dB or 0 dB (as per $o_{prep}$, set uniformly in each simulation run). The execution of CHO (controlled by $o_{exec}$) occurs when target cell's received power is 3 dB better than that of the source cell. This value of $o_{exec}$ is typically used for handovers in NR, including non-conditional mobility. It ensures low level of handover failures (HOFs) and ping-pongs (PP). The latter occurs when the UE returns to the source cell shortly after completing the handover at the target cell (i.e. usually within 1 second).

TABLE II: SIMULATION PARAMETERS

| Parameter | Value |
|---|---|
| Inter-Site Distance (ISD) | 200 m |
| Deployment | 7 sites, hexagonal cells |
| BS Tx power | 30 dBm |
| Channel Model | Urban Micro (UMi), compliant with 3GPP Technical Report 38.901 |
| Number of UEs | 420 |
| Carrier frequency | 28 GHz |
| Simulation time | 30 seconds |
| SINR outage limit | -8 dB |
| UE mobility model | Random waypoint |
| CHO preparation offset ($o_{prep}$) | 10 dB, 3 dB, 0 dB |
| CHO execution offset ($o_{exec}$) | 3 dB |
| Maximum number of prepared CHO candidate cells ($M$) | 1, 2, 3 |
| Number of beams per CHO candidate with CFRA resources | 1 |
| CFRA Threshold ($Thr_{CFRA}$) | (-97 dBm, -70 dBm) with 3 dB step |
| CFRA update duration ($T_{update}$) | 30 ms |
| Number of RA preamble retransmissions ($n$) | 0, 1, 2, 3, 4 |

The simulations have been run for different number of maximum candidate CHO cells (denoted by $M$), with three being the highest investigated value. Verifying larger number of CHO candidate cells is not justified as each additional cell prepared early in advance increases the burden on the network side to keep dedicated cell-specific and beam-specific radio resources. As claimed in [5], the main benefit of CHO is due to separation of handover phases, not due to preparing a vast set of candidate cells, which confirms our choice of the investigated range of parameter $M$.

### B. KPIs and Investigated Metrics

The KPIs which have been derived for assessing the performance are as follows:
- CBRA rate ($R_{CBRA}$) calculated as the percentage of contention-based RA attempts over all RA attempts
- Handover Failures ($N_{HOF}$) calculated as the number of failed handovers normalized over the number of UEs and time
- Number of CFRA updates ($N_{update}$) normalized over the number of UEs and time
- Average HO delay ($D_{HO}$) for two cases: with and without CFRA updating procedure (denoted by $D_{HO}^{with}$ and $D_{HO}^{w/o}$, respectively) calculated using the average time for completing CFRA-based HO, considering the preamble retransmissions, the time needed for CBRA (introduced in section II.A) and involving the $R_{CBRA}$ obtained for each $Thr_{CFRA}$.

CBRA rate ($R_{CBRA}$) was derived using the following formula

$$R_{CBRA} = \frac{N_{CBRA}}{N_{CFRA}+N_{CBRA}} \cdot 100\% \qquad (3)$$

where $N_{CBRA}$ and $N_{CFRA}$ represent the total number of CBRA and CFRA handovers, respectively. HOF is recorded when the UE fails to successfully complete the CHO (i.e. the timer T304 controlling the CHO execution has been started but HO was eventually unsuccessful [6]). Average HO delay ($D_{HO}$) is calculated separately for the cases with and without CFRA resources updating, as per the following equation

$$D_{HO} = R_{CBRA}(HO_{CFRA} + n \cdot D_{CBRA}^{avg}) + (1 - R_{CBRA})HO_{CFRA} \qquad (4)$$

where $HO_{CFRA}$ is the time needed to complete handover using CFRA. In our study it is equal to 80 ms, as estimated in [12]. $D_{CBRA}^{avg}$ represents the time it takes to complete the CBRA and is equal to the sum of all components in Table I, while $n$ is the number of RA preamble retransmissions needed, e.g. due to a collision with other user.

### C. Simulation Results

First set of results in Fig.4 depicts the $R_{CBRA}$ for different $Thr_{CFRA}$ (simulated range of $Thr_{CFRA}$ described in Table II). The curves have been obtained for multiple CHO preparation offsets ($o_{prep}$) and for a maximum number of three prepared CHO candidate cells ($M$). As can be noticed, updating CFRA resources in line with the condition (2) brings desirable reduction of $R_{CBRA}$ in all investigated scenarios. The curves representing cases where CFRA resource updating was applied are nearly the same for the entire evaluated range of $Thr_{CFRA}$. It implies that even for $o_{prep}$ of 0 dB updating beam-specific CFRA resources was feasible in most of the cases before the CHO execution occurred ($o_{exec}$ set to 3 dB). The improvement is most evident for the $o_{prep}$ of 10 dB where $R_{CBRA}$ for the case without CFRA updating was the highest for all investigated $Thr_{CFRA}$ values (purple curve in Fig. 4). This is due to the early preparation (when source cell remains 10 dB stronger than candidate target cell) which often leads to the prepared beam at the time of handover execution not exceeding $Thr_{CFRA}$. This effect is stronger for higher values of $Thr_{CFRA}$ while using such high settings for this parameter is beneficial for ensuring CFRA resources are only used on sufficiently good beams.

The difference between the curves for cases with and without CFRA updating is negligible for the lowest values of $Thr_{CFRA}$ (below -88 dBm) but increases for higher $Thr_{CFRA}$ and reaches 13% for RSRP threshold of -79 dBm and -76 dBm (for $o_{prep}$ of 10 dB). For lower values of $o_{prep}$ the gains are smaller but still can reach 7 - 8% (e.g. for -79 dBm and $o_{prep}$ of 3 dB).

The improvement in $R_{CBRA}$ comes at certain expense – increased signalling between the UE and the source BS for reporting measurements and providing new CFRA resource assignments. Fig. 5 illustrates how many successful CFRA updates occurred per UE per minute when different number of CHO candidate cells ($M$) was prepared. As can be noticed, the number of CFRA updates per UE does not decrease linearly with the reduced number of prepared CHO candidate cells.

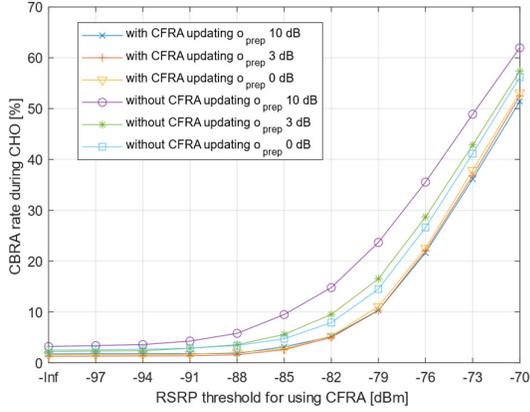

Fig. 4. CBRA rate against RSRP beam threshold for CFRA-based handover when up to three CHO candidate cells are prepared.

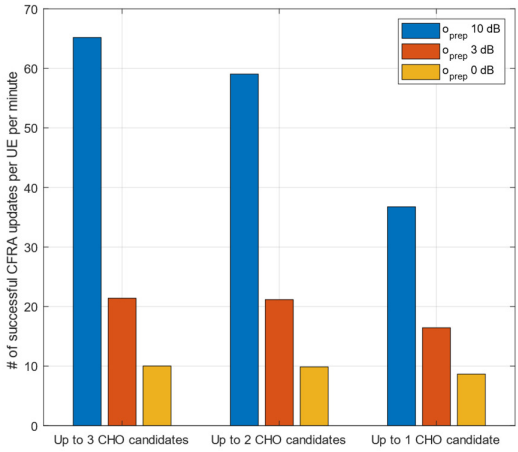

Fig. 5. Number of successful CFRA updates ($N_{update}$) for different $o_{prep}$ and maximum number of prepared CHO cells (M).

This is due to updating the CFRA beam assignment only if currently prepared beam is not the best per each candidate cell, whereas a single measurement report may comprise results for all prepared cells from set C, some of which will not obtain new CFRA assignments due to not fulfilling the condition in (2). Similar explanation can be provided for no significant difference between the results for up to two CHO candidates and up to three CHO candidates – in most cases the CFRA resources for a single prepared cell $c_i$ will be subject to reassignment, while the beams for the remainder of prepared cells do not meet the (2).

As expected, the number of updates is the largest for the $o_{prep}$ of 10 dB, as the UE is configured with CHO early, including beam-specific CFRA resource assignments, so there is significant time between CHO preparation and execution for radio signal fluctuations and measurement reporting. In addition, the initial assignments are likely to become obsolete if the candidate target cell was configured to the UE when it was still 10 dB worse than serving cell.

Yellow bars in Fig. 5, representing $o_{prep}$ of 0 dB, do not vary significantly for different number of prepared CHO candidates, as the relatively late preparation (i.e. close to CHO execution) usually leads to choosing optimal beam, which shortly later can be used for actual CFRA-based handover.

As shown above, the updating of CFRA resources is confirmed to increase the ratio of CFRA-based HO execution. However, another desirable goal is to reduce the number of failed HOs (HOFs) via providing CFRA resources to best beam at the time of CHO execution. Fig. 6 depicts the number of HOFs for different $Thr_{CFRA}$ and different number of CHO candidates (M). All curves shown in Fig. 6 represent the scenario of early preparation ($o_{prep}$ of 10 dB). It can be noticed that regardless of the number of prepared CHO candidates (M) the number of HOFs is higher in case CFRA resource updating is not used compared to the cases where such updating is applied. This is especially visible for lower values of $Thr_{CFRA}$, while the results become very similar starting from –85 dBm. It implies updating CFRA resources before HO can lead to further reduction of the number of HOFs the UEs encounter, at least in case of early preparation and when the BS configures the $Thr_{CFRA}$ to lower values. Such settings of $Thr_{CFRA}$ can be considered to maximize the use of CFRA at the risk of increased HOF rate. The latter can be reduced by applying CFRA updating method, studied in this paper.

It is worth noting that the number of HOFs presented in Fig. 6 is considerably low and evaluated handover attempts have led to a failure in less than 1% of the simulated cases. This proves CHO alone is a technique that already ensures high robustness, while remaining HOFs can be eliminated in some scenarios, e.g. by applying CFRA updating.

Finally, in Fig. 7 we present the curves showing how the average HO delay ($D_{HO}$) changes when CFRA resource updating procedure is applied. The results were derived using (4) and considering varying number of RA preamble retransmissions (n), for the maximum of three CHO candidate cells (M) and $o_{prep}$ of 10 dB. For the lowest values of $Thr_{CFRA}$ (i.e. not exceeding -88 dBm) $D_{HO}$ is not larger than 85 ms and does not differ much from $HO_{CFRA}$, representing the scenario where most handovers would be executed using CFRA. For larger $Thr_{CFRA}$ the handover delay ($D_{HO}$) clearly increases and reaches up to 128 ms for the case with no CFRA resource updating and four RA preamble retransmissions. The average HO duration can be reduced to approximately 120 ms, when CFRA resource updating is applied and four RA preamble retransmissions are needed. Fig. 7 confirms that using CFRA resource updating procedure is justified in the scenarios where preamble collisions are likely and when the network sets relatively high $Thr_{CFRA}$ (i.e. larger than -85 dBm), to ensure CFRA resources are used only when associated beam is sufficiently strong.

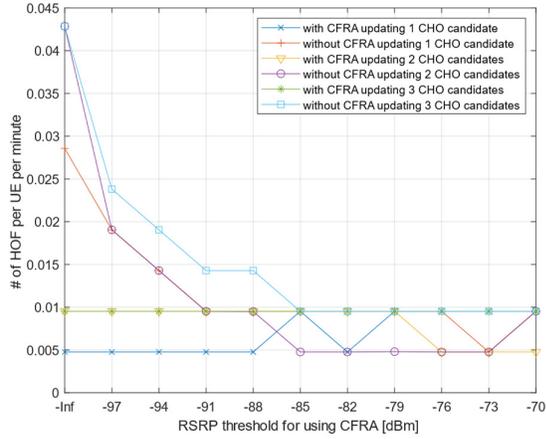

Fig. 6. Number of HOFs ($N_{HOF}$) for different CFRA thresholds and number of CHO candidates ($o_{prep}$ 10 dB).

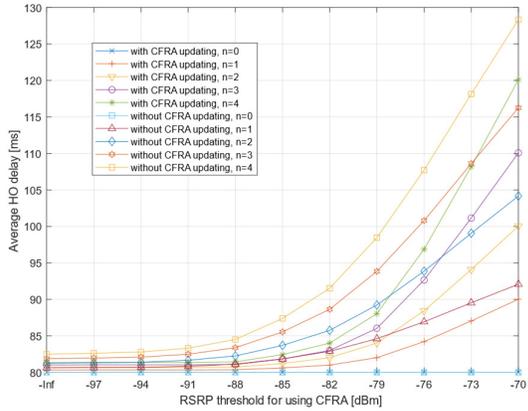

Fig. 7. Average HO delay for different CFRA thresholds, maximum three CHO candidate cells and $o_{prep}$ 10 dB.

## IV. CONCLUSION

In this article we have analyzed the contention-based and contention-free random access during conditional handover. We have identified the issue of CFRA resources becoming unusable in CHO as the time between handover preparation and execution is much longer when compared to non-conditional mobility. To mitigate this CHO deficiency, we have proposed a method wherein the UE can initiate additional, beam-level measurement reporting and the BS may update the beam-specific CFRA resource allocations. As shown in the simulation study, our proposed solution allows to reduce the CBRA rate by up to 13% for typical values of CFRA application thresholds. This leads to a decrease of the average HO delay by approximately 10 ms. We have shown that CFRA resource updating can help in further reduction of HOF ratio, at least for some values of $Thr_{CFRA}$. For completeness, we have also presented at what expense these gains can be achieved, i.e. how much the radio signaling between the UE and the source BS increases due to performing CFRA resource reassignment procedure.

Our future work will concentrate on verifying whether there are additional gains if the source BS is provided with a pool of CFRA resources in advance and subsequent coordination with the target BS is not necessary. We will also focus on optimizing the CFRA resource updating procedure by selectively choosing the CHO candidate cells to be subject to such reassignment process. This shall lead to decreased radio signaling, while not impacting much the overall mobility performance. One of the considered approaches could be to derive cell-specific probabilities reflecting how likely each of these cells will be accessed and then apply this metric in the measurement report triggering process.